\documentclass{aa}
\usepackage[utf8]{inputenc}
\usepackage{graphicx}
\graphicspath{{graphs/}} 
\usepackage{xcolor}
\begin{document}

\title{HD\,38858: a solar type star with a $\sim$10.8 yr activity cycle \thanks{Based on observations made with ESO Telescopes at the La Silla Paranal Observatory under programmes: 183.C-0972(A), 072.C-0488(E), 192.C-0852(A), 091.C-0936(A) and 198.C-0836(A).} }

\subtitle{Searching for Balmer and metallic line variations.}
\author{M. Flores\thanks{Visiting Astronomer, Complejo Astron\'omico El Leoncito operated under
agreement between the Consejo Nacional de Investigaciones Cient\'ificas y T\'ecnicas de la República Argentina, the National Universities of La Plata,
C\'ordoba, San Juan.}
\inst{1,3,6}
    \and
J. F. Gonz\'alez
\inst{1,3,6}
\and
M. Jaque Arancibia
\inst{1,6,7}
\and
C. Saffe\inst{1,3,6}
\and
A. Buccino\inst{2,4,6}
\and
F. M. L\'opez\inst{1,3,6}
\and
R.V Ibañez Bustos\inst{2,6}
\and
P. Miquelarena\inst{3}
}

\institute{Instituto de Ciencias Astron\'omicas, de la Tierra y del Espacio (ICATE), Espa\~na Sur 1512,
    CC 49, 5400 San Juan, Argentina.
    \email{matiasflorestrivigno,jfgonzalez,mjaque@conicet.gov.ar}
    \and
Instituto de Astronom\'ia y F\'isica del Espacio (IAFE), Buenos Aires, Argentina.
\and
Facultad de Ciencias Exactas, F\'isicas y Naturales, Universidad Nacional de San Juan, San Juan, Argentina.
\and
Departamento de F\'isica, Facultad de Ciencias Exactas y Naturales, Universidad de Buenos Aires, Buenos Aires, Argentina.
\and
Observatorio Astron\'omico de C\'ordoba (OAC), Laprida 854, X5000BGR, C\'ordoba, Argentina
\and
Consejo Nacional de Investigaciones Cient\'ificas y T\'ecnicas (CONICET), Argentina.
\and
Departamento de F\'isica y Astronom\'ia, Universidad de La Serena, Av. Cisternas 1200, La Serena, Chile
}
\date{February 2018}

\abstract
{The detection of chromospheric activity cycles in solar-analogue and twin stars, can be used to put in context the solar cycle. However, there is a relatively low percentage of these stars with activity cycles detected. It is well known that the cores of \ion{Ca}{ii} H\&K lines are modulated by stellar activity. The behaviour of Balmer and other optical lines with stellar activity is not yet completely understood.}
{To search for \ion{Ca}{ii} H\&K, Balmer and \ion{Fe}{ii} line variations modulated by the stellar activity. In particular, we apply a novel strategy to detect possible shape variations in the H$\alpha$ line.}
{We analyse activity signatures in HD 38858 by using HARPS and CASLEO spectra obtained between 2003 and 2017. We calculate the Mount Wilson index (S$_{MW}$), log($\mathrm{R}'_\mathrm{HK}$) and the statistical moments of \ion{Ca}{ii} H\&K, Balmer and other optical lines. We search for periodicities by using the generalized Lomb-Scargle periodogram.}
{We detect a long-term activity cycle of 10.8 yr in \ion{Ca}{ii} H\&K and H$\alpha$, being HD 38858 one of the most similar stars to the Sun for which a cycle has been detected. In contrast, this cycle is marginally detected in the \ion{Fe}{ii} lines. Also, we detect a noticeable radial velocity variation, which seems to be produced by stellar activity.}
{HD 38858 is the second solar-analogue star where we find a clear activity cycle which is replicated by Balmer lines. Spectral indexes based on the shape of H$\alpha$ seems to be more realiable than the fluxes in the same line to detect activity variations. On the other hand, the cyclic modulation detected gives place to a radial velocity variation, previously associated to a super-Earth planet. Finally, due to the similarity of HD 38858 with the Sun, we recommend to continue monitoring this star.}
{}


\maketitle

\section{Introduction}

\citet{1978ApJ...226..379W} showed for the first time that solar-type stars could evidence long-term chromospheric variations in their \ion{Ca}{ii} H\&K lines. This was followed by a number of studies, mainly the HK project at the Mount Wilson Observatory \citep[e.g.][]{1978PASP...90..267V,1991ApJS...76..383D,1995ApJ...438..269B,1996AJ....111..439H}. Different works also showed that stellar activity is related to important variables such as rotation, differential rotation, and the stellar age \citep[e.g.][]{1972ApJ...171..565S,1993PhDT.........3D,2008ApJ...687.1264M}. These studies established the basis of the current knowledge of activity for the stars in the solar neighborhood.

Special attention deserve activity studies of stars physically similar to our Sun, helping to put its $\sim$11 yr cycle in context. For instance, the solar-twin star 18 Sco presents a cycle of 7.1 yr \citep{2007AJ....133..862H}, while the solar-analog HD\,30495 shows short-period variations together with a long cycle of $\sim$12 yr \citep{2015ApJ...812...12E}. Both stars 
 are younger than the Sun, being 3.8 $\pm$0.5 Gyr the age of 18 Sco \citep{2015A&A...579A..52N} and 0.97 $\pm$ 0.12 Gyr that of HD\,30495 \citep{2015ApJ...812...12E}. In addition, we found a cycle of only $\sim$5.1 yr in the solar analog HD\,45184 \citep{2016A&A...589A.135F}, while in the $\zeta$ Ret binary system the components show erratic ($\zeta^{1}$ Ret) and cyclic activity patterns ($\zeta^{2}\sim$10 yr) simultaneously \citep{2018MNRAS.476.2751F}, being both stars similar to the Sun \citep{2016A&A...588A..81S}.

We have an ongoing program currently monitoring the stellar activity in a sample of solar-analogue and solar-twin stars. For our studies, we mainly use the extensive data base of HARPS (High Accuracy Radial velocity Planet Searcher) spectra, which occasionally are complemented with CASLEO observations. Some recent findings of our new programme can be found in \citet{2016A&A...589A.135F,2017MNRAS.464.4299F,2018MNRAS.476.2751F}.

Our sample includes star the HD\,38858 (=HIP\,27435). This nearby object is located at $\sim$15 pc \citep{2016A&A...595A...2G},
 presents a $B-V$ color of 0.64 \citep{2002yCat.2237....0D}, and spectral type G2 V \citep{2003AJ....126.2048G}. In addition, the following atmospheric parameters have been reported: $T_\mathrm{eff} =$ 5733  $\pm$12 K, $\log g$ $=$ 4.51 $\pm$0.01, [Fe/H] $\sim$ -0.22 $\pm$ 0.01, and v$_\mathrm{turb} =$ 0.94 $\pm$0.02 km\,s$^{-1}$ \citep{2017A&A...606A..94D}. 
Although the low metallicity value of HD 38858 ([Fe/H] $\sim$ -0.22 $\pm$ 0.01) would leave it out of some solar-analogue star definitions \citep[e.g.][]{2010A&A...521A..33R}, it is considered, even following different criteria, as a solar-analogue star by several literature works \citep[e.g.][]{2009AJ....138..312H,2010ApJ...720.1592G,2014A&A...562A..92D}. In addition, it is important to point out that the ``top ten of solar analogs'' from \citet{2004A&A...418.1089S} is composed for stars whose metallicities range from -0.23 to +0.11 dex.

At this moment, the HARPS monitoring of this star provides a large data set, enough to make a detailed long-term stellar activity study of HD\,38858. Moreover, due to the solar-analogue nature of this star, it could be used for a direct comparison with our Sun. In this way, HD\,38858 constitutes an excellent laboratory to perform a comparative study of their chromospheric patterns (e.g. mean activity level, cycle length, cycle amplitude, etc.), which has been done for few stars, such as 18 Sco \citep{2007AJ....133..862H} and HD\,45184 \citep{2016A&A...589A.135F}.

For the solar case, \citet{1982ApJ...252..375L} first revealed the spectral variation of several Fraunhofer lines along its $\sim$11 yr activity cycle (well tracked by the \ion{Ca}{ii} H\&K lines). Subsequently, \citet{2007ApJ...657.1137L} showed that the \ion{Fe}{} lines did not show the expected modulation with the \ion{Ca}{ii} H\&K lines. For the stellar case, the possible cyclic modulation between \ion{Ca}{ii} H\&K with both H$\alpha$ and metallic-line variations is poorly known. \citet{2014A&A...566A..66G} studied the correlation between the flux of \ion{Ca}{ii} H\&K and H$\alpha$ (through the $I_{H\alpha}$ index) lines in an stellar sample of 271 FGK stars, finding that 23\% out of 271 stars are correlated. In particular, the solar analogue star HD 45184 is included in the group of stars that do not show a correlation between these fluxes. However, we detected variations in Balmer and Fe lines in this star \citep{2016A&A...589A.135F}, modulated by its $\sim$5.1 yr chromospheric activity cycle. The behaviour of the lines in this object is different to that observed in the Sun, in spite of their similar stellar parameters. Another solar-analogue star which does not show correlation is HD 38858 (see table B.2. of \citealt{2014A&A...566A..66G}). Then, similar to the case of HD 45184, studying the variability of the H$\alpha$ line in HD 38858, can be used to test the \ion{Ca}{ii} H\&K-$I_{H\alpha}$ correlation.

These puzzling patterns show that there is a need of long-term activity studies of stars similar to our Sun, which are  essential to a complete  understanding of the observed phenomena.

Both main techniques for detection of extrasolar planets, i.e., transits and radial velocities, can be affected by stellar activity \citep[e.g.][]{2008NewAR..52..154S,2011PhDT........60M}. In particular, radial velocity (RV) could vary due to the presence of active regions, which hinder the convection pattern \citep{1988ApJ...331..902C,1985srv..conf..311D,1992ESOC...40...55D} and gives place to a correlation between stellar activity and RV \citep[e.g.][]{1992ESOC...40...55D,2012A&A...541A...9G,2011A&A...535A..55D}. In this way, an activity cycle could introduce a periodic signal into the RV, which can be easily confused with an exoplanet \citep[e.g.][]{2010A&A...520A..79M,2011A&A...535A..55D,2012A&A...541A...9G,2014A&A...567A..48C}. In fact, \citet{2011arXiv1109.2497M} announced through a preprint the discovery of a radial super-Earth planet (minimum mass of $30.55 M_{\oplus}$) orbiting HD\,38858 with a period of 407.15 days, although this detection was not confirmed later. Recently, using a more extensive data set, \citet{2015MNRAS.449.3121K} pointed out that the signal associated to this supposed planet is an alias of a stellar activity cycle (with a period of 2930 d). However, through these additional data, the authors found another planetary signal with a period of 198$\pm$1 d (0.64 AU) and minimum mass of 12$\pm$2 $M_{\oplus}$. Nevertheles, it should be stressed that the corresponding analysis has not been published yet.

Although stellar activity can be a source of noise in precise RV measurements, it can be corrected if activity is simultaneously measured using indexes, such as \ion{Ca}{ii} H\&K and H$\alpha$ lines. Hence, the study of the activity proxies and their relation with RV is crucial to diminish the activity effects on RV measurements \citep{2010A&A...520A..79M,2011arXiv1107.5325L,2014A&A...566A..66G}. In other words, it could provide valuable information to distinguish between stellar and planetary signals. Fortunately, the long-term database available of the solar-analogue star HD 38858 allows us to carry out this type of study.

This work is organized as follows. In §2 we describe the observations and data reduction, while in §3 we describe our
analysis of chromospheric activity and in §4 we present our discussion and conclusions.

\section{Observations and data reduction}

Our study is based on spectra of HD\,38858 downloaded from the European Southern Observatory (ESO) 
archive\footnote{\url{http://archive.eso.org/wdb/wdb/adp/phase3_spectral/form?phase3_collection=HARPS}} \citep{2003Msngr.114...20M}.  
The spectra were acquired with the HARPS spectrograph, attached to the La Silla 3.6 m (ESO) telescope. This spectrograph is fed by a pair of fibres with an aperture of 1 arcsec on the sky, resulting in a resolving power of $\sim$115 000\footnote{{\url{http://www.eso.org/sci/facilities/lasilla/instruments/harps/overview.html}}}. The observations were taken between 2003 to 2017 and have been automatically processed by the HARPS pipeline\footnote{\url{http://www.eso.org/sci/facilities/lasilla/instruments/harps/doc.html}}, covering a spectral range between 3782--6913 {\AA}. After discarding a few low signal-to-noise (S/N) observations, we obtained a total of 237 spectra with a S/N ranging from 100 to 350 at 6075 {\AA}. Then, these spectra were carefully normalized, and cleaned from cosmic rays and telluric features using IRAF\footnote{IRAF is distributed by the National Optical Astronomical Observatories, which is operated by the Association of Universities for Research in Astronomy, Inc., under a cooperative agreement with the National Science Foundation.} routines, as in our previous work \citep{2016A&A...589A.135F}.

We complement our analysis with observations obtained at the Complejo Astronómico El Leoncito (CASLEO, San Juan -- Argentina). We used the REOSC spectrograph mounted at the 2.15 m telescope between the years 2012 and 2014. These mid-resolution echelle spectra  (R$\sim$13 000) ranging from 3890 to 6690 {\AA}, were reduced using standard IRAF routines.

In order to compute the standard Mount Wilson index (S$_{MW}$), we integrated the flux in two
windows centred at the cores of the \ion{Ca}{} lines, weighted with triangular profiles of 1.09 {\AA} full width at half-maximum (FWHM), and 
computed the ratio of these fluxes to the mean continuum flux, integrated in two passbands of 20 {\AA} width centred at 3891 and 4001 {\AA}.
Finally, the S$_{MW}$ indexes from both HARPS and CASLEO spectra, were obtained by using the calibration procedures explained in \citet{2011arXiv1107.5325L} and \citet{2007A&A...469..309C}, respectively \citep[see][for details]{2017MNRAS.464.4299F}.

\section{Activity cycle}
\subsection{Chromospheric emission at \ion{Ca}{ii} H\&K lines}

\begin{figure}[htb!]
\centering
\includegraphics[width=\hsize]{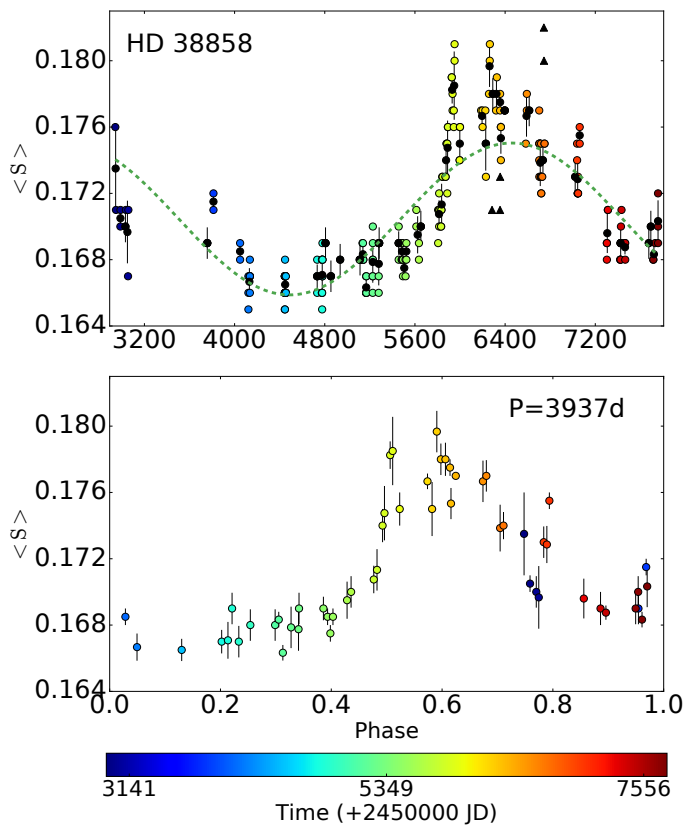}
\caption{Upper panel: Individual S$_{MW}$ measurements (coloured circles) and seasonal means (black full circles) from HARPS observations. The black triangles correspond to CASLEO data. 
The dotted green line shows a harmonic curve with the cycle calculated in this work. Lower panel: Mt Wilson index phased with a 10.8 yr period.}
\label{plot.0}
\end{figure}

\begin{figure}[htb!]
\centering
\includegraphics[width=\hsize]{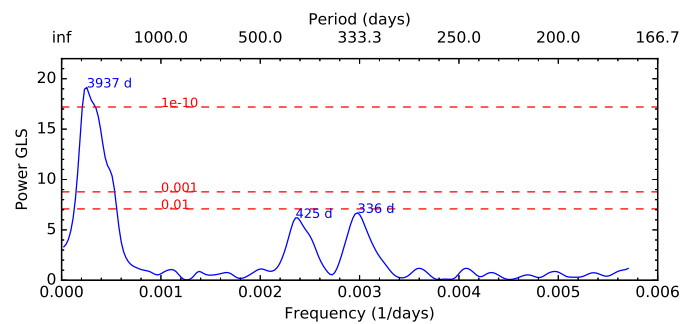}
\caption{GLS periodogram for the Mount Wilson indexes data plotted in Figure \ref{plot.0} (upper panel).}
\label{plot.1}
\end{figure}

In Figure \ref{plot.0} (upper panel) we show the time series of the 
S$_{MW}$ for HD\,38858. Here, we included all data, i.e. HARPS and CASLEO observations and also the corresponding monthly average values. These averaged data allow us to remove short-timescale variations caused by the transit of individual active regions (spots and plages). This same procedure has been applied in previous works in the literature \citep[e.g.][]{1995ApJ...438..269B,2010ApJ...723L.213M,2018MNRAS.476.2751F}. The error bars of HARPS data correspond to the standard deviation of the mean. For those bins with only one measurement, we adopted the typical dispersion of other bins. Then, we derived a long-term activity period of 3940 $\pm$ 290 d ($\sim$ 10.8 yr) by applying the generalized Lomb-Scargle periodogram \citep[thereafter GLS,][]{2009A&A...496..577Z} to the seasonal means (Figure~\ref{plot.1}). The false-alarm probability (thereafter FAP) of the main resulting period of 3.23 $\times 10^{-13}$ was calculated following \citet{2009A&A...496..577Z} (see their equation 24 for details). It can be noted that together with the more significant peak there are two additional, less significant ones at $\sim$425 $\pm$ 8 d (FAP $\sim$3.08 $\times 10^{-2}$) and $\sim$336 $\pm$ 6 d (FAP $\sim$1.65 $\times 10^{-2}$). However, due to we adopted a cut-off in FAP of 0.1 per cent (0.001) for reliable periodicities, both periods (likely aliases of the long-term cycle) have not been taken into account. We stress that, according to \citet{1995ApJ...438..269B}, our period FAP$\sim$3.23 $\times 10^{-13}$ would be classified as excellent (FAP $\leq$ $\times 10^{-9}$).

It is important to mention that HD\,38858 not only has an activity cycle similar to our Sun ($\sim$11 yr), but also its mean chromospheric activity level log($\mathrm{R}'_\mathrm{HK}$) = -4.94 is very similar to the solar values published by \citet{2007AJ....133..862H} and \citet{2008ApJ...687.1264M} (-4.94 and -4.91, respectively). This activity index is obtained after substracting the photospheric contribution, according to the \citet{1984ApJ...279..763N} calibration. It is also worth mentioning that, according to \citet{1995ApJ...438..269B}, HD 38858 is simultaneously a near flat star ($\sigma_{\mathrm{S}}$/$\overline{\mathrm{S}}\approx 1.5\%-2\%$) and a cycling one (FAP $\leq 10^{-2}$). Furthermore, in Figure \ref{plot.0}, we observe that $S_{MW}$-cycle shape is similar to the sunspot cycles \citep{1994SoPh..151..177H}. It presents a sudden rise toward its maximum and a slower decay to the minimum. \cite{2011SoPh..273..231D} proposed a modified Gaussian distribution  to fit and predict the solar cycles. This formula was recently corrected by \cite{2017ApJ...835...25E} with an offset to the expression:

\begin{equation}
f=A \exp \left(-\frac{(t-t_m)^2}{2B^2[1+\alpha(t-t_m)]^2}\right)  +f_{min}
\label{eq.shape}
\end{equation}
where $A$ represents the maximum value, $t_m$ the time when the cycle reach the maximum if $t$ is computed from the minimum, $B$ is the width of Gaussian ``bell'' and $\alpha$ is an assymetry factor. Although Eq. \ref{eq.shape} can model successfully several solar cycles, it cannot reproduce the double peaks present in recent cycles.
\begin{table}
\caption{Fit parameters from Eq. \ref{eq.shape} for HD\,38858.}             
\label{table:1}      
\centering

\begin{tabular}{ll}      
\hline\hline                
Parameter & Value\\ 
\hline                     
$A$ & 0.0101$\pm$ 0.0001\\    
$B$ & 1.44 $\pm$  0.04 yr\\
$\alpha$ & 0.1052 $\pm$ 0.019 yr$^{-2}$\\
$t_m$ & 2013.018 $\pm$ 0.063 \\
$f_{min}$ & 0.16709 $\pm$ 0.0002\\
\hline                                 
\end{tabular}
\end{table} 

 In particular, \cite{2017ApJ...835...25E} applied this formula to fit the Mount Wilson indexes obtained for the Sun during its Cycle 23. Following the numerical method described in this work, we fit Eq.\ref{eq.shape} to the HD\,38858 stellar cycle and obtained the parameters listed in Table \ref{table:1} with a Pearson correlation coefficient of 0.90.

\begin{figure}[h!]
\includegraphics[width=\hsize]{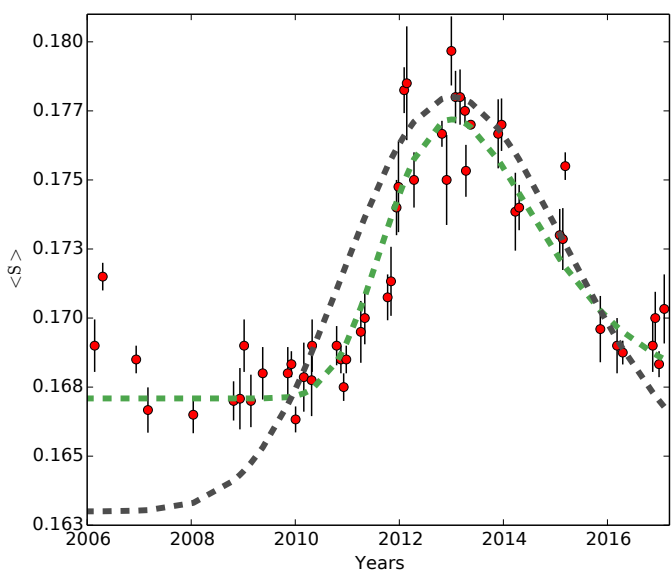}
\caption{Time series of the Mt Wilson indexes for HD\,38858 (full red circles). The dashed green line corresponds to the fitted cycle, while the dashed gray line corresponds to the cycle derived for the 23th. solar cycle.}
\label{fig.cycleshape}
\end{figure}

In Fig. \ref{fig.cycleshape} we show the mean Mount Wilson series (full red circles) and the corresponding cycle fit obtained for HD\,38858 (dashed green line), together with the cycle derived by \cite{2017ApJ...835...25E} for the 23th. solar cycle (dashed gray line). Thus, if we compare the obtained parameters for the fitted stellar cycle with the solar ones ($A_{\odot}=0.0150$, $B_{\odot}=2.154$ and $\alpha_{\odot}=0.0343$), we observe that they are slightly different. However, the shape proposed in Eq. \ref{eq.shape} fit successfully the stellar cycle.

The star HD 38858 has an activity period of $\sim$ 10.8 $\pm$ 0.8 yr. On the other hand, the period of the solar activity cycle ranges between 9 and 13 yr, 
with an average of 11 yr and a standard deviation of about 1.16 yr \citep{2010LRSP....7....1H}. Then, it is difficult to know if 
a metallicity difference of 0.22 dex between both stars plays a role in the length of the periods. Up to now, there is no clear observational evidence for the relation between the cycle period and metallicity. For instance, \citet{2017ApJ...845...79B} analyzed the cycle periods of 35 FGK stars, and did not find a clear relation with metallicity. Thus, studying a larger sample of solar-analogues and solar-twin stars with measured activity cycles and metallicities very different to the Sun, could help to clarify this issue.

\subsection{An activity mimicking case or a super-Earth planet orbiting HD\,38858?}

\begin{figure}
\centering
\includegraphics[width=\hsize]{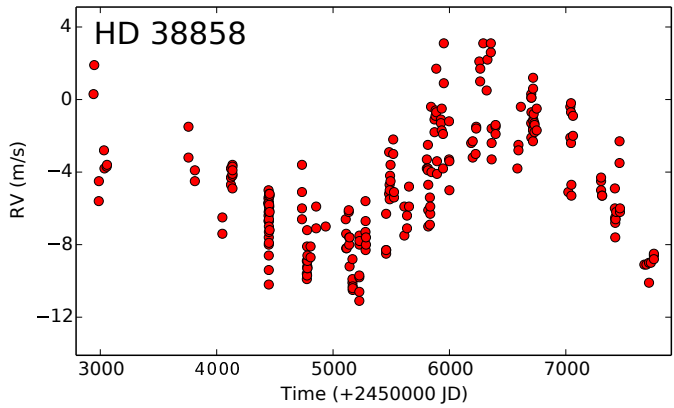}
\caption{Time series of RV measurements obtained from HARPS spectra.}
\label{plot.2}
\end{figure}

\begin{figure}
\centering
\includegraphics[width=\hsize]{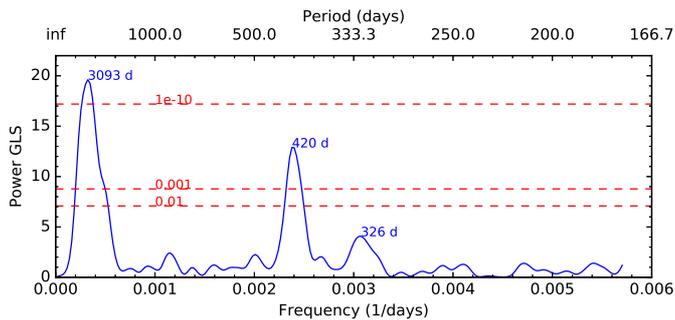}
\caption{GLS periodogram for radial velocities plotted in Figure \ref{plot.2}.}
\label{plot.3}
\end{figure}

\begin{figure}
\centering
\includegraphics[width=\hsize]{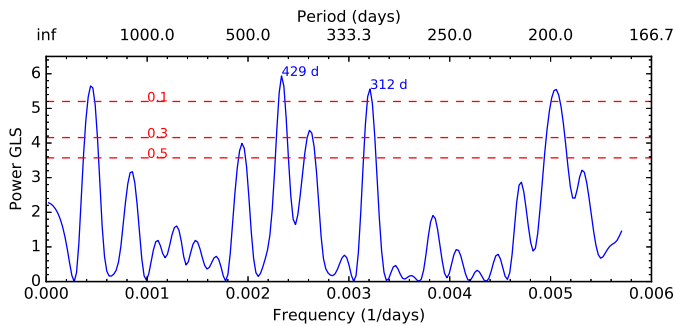}
\caption{GLS periodogram after subtracting the $\sim$ 3100 d period.}
\label{plot.2.}
\end{figure}

In Figure~\ref{plot.2} we plot the time series of the RV measurements for all HARPS spectra. A first glance of this figure reveals the presence of a clear modulation, which also seems very similar to the activity one (see Figure~\ref{plot.0}). As we previously mentioned, this RV variation might be produced by the stellar activity and not by an extrasolar planet. To test this possibility, we calculate the GLS periodogram for these data (Figure~\ref{plot.3}). As a result, we detect a significant period of $\sim$ 3100 $\pm$200 d with a FAP of 6.47 $\times 10^{-14}$ and two short less significant periods of $\sim$420 $\pm$ 4 d (FAP $\sim$1.28 $\times 10^{-6}$) and $\sim$326 $\pm$ 8 d (FAP $\sim$3.24 $\times 10^{-1}$), respectively. We note that these three peaks are close to those values obtained from the stellar activity analysis (see Figure~\ref{plot.1}). Then, in order to rule out possible alias peaks of the long-term signal (i.e. $\sim$ 3100 d), we subtracted it and recalculated once again the corresponding GLS periodogram, according to \citet{2016A&A...595A..12S} and \citet{2018MNRAS.476.2751F}. As a result, in Figure \ref{plot.2.} we can observe that both less significant periods are not present now, which indicates that these peaks would correspond to alias.  

\begin{figure}
\hskip -0.3cm
\includegraphics[width=\hsize]{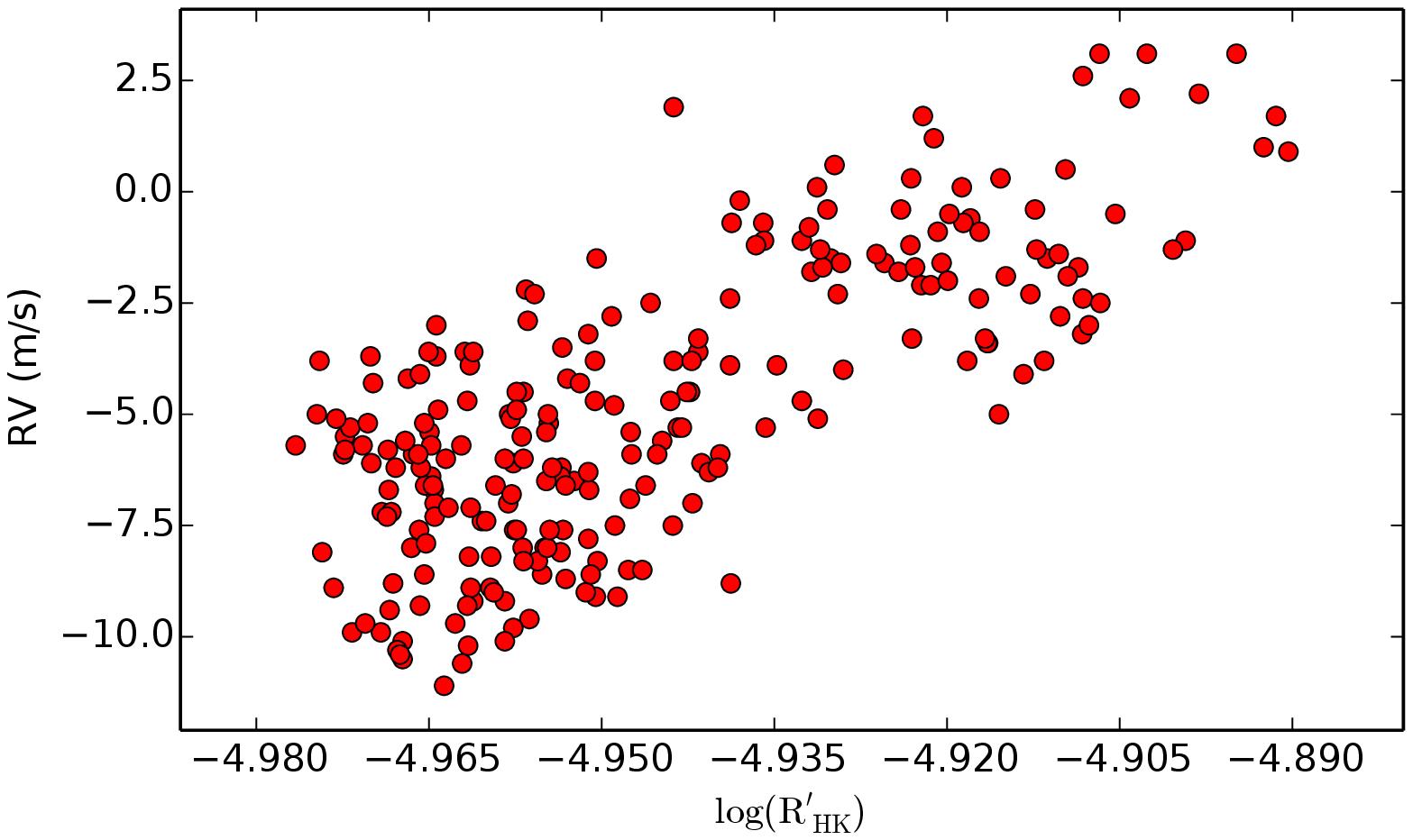}
\caption{RV vs. log($\mathrm{R}'_\mathrm{HK}$) for HD\,38858.}
\label{plot.4}
\end{figure}

In addition, in Figure~\ref{plot.4} we plotted RV vs. log($\mathrm{R}'_\mathrm{HK}$). We applied a Bayesian analysis to these data for asses the presence of a possible correlation. Using the \textit{python} code provided by \cite{Figueira16}, we estimated the posterior probability distribution of the correlation coefficient $\rho$. From this test, we obtained  a correlation coefficient of 0.744 $\pm$ 0.029 with a 95\% credible interval between 0.688 and 0.800. Thus, the similarity between the time series together with the GLS periodogram analysis, as well as, the RV--log($\mathrm{R}'_\mathrm{HK}$) trend, suggest that the signal attributed to HD\,38858b could be related to stellar activity and probably not to a super-Earth as claimed by \citet{2011arXiv1109.2497M} and so considered in subsequent statistical works \citep[e.g.][]{2014A&A...562A..92D,2015A&A...580A..24D,2016MNRAS.462.1563M}. This conclusion was first achieved by \citet{2015MNRAS.449.3121K}. However, until now the corresponding analysis that supports this claim has not been published.

\subsection{Searching for Balmer and metallic lines variations}

We applied different techniques to search for small spectral variations in hydrogen and metallic lines throughout the activity cycle.
As a prior step, we made a general evaluation of noise in  our spectroscopic material in order to
estimate the level above which eventual variations can be considered reliable.
This was motivated by the detection of non-random modulations in residual spectra (individual-minus-average), that might impose a minimum threshold for long-term variability detection.
The shape of these systematic patterns suggests that they are related to normalization, echelle order tracing, order merging, and wavelength calibration problems. 

In order to estimate the typical noise level we calculate the RMS of residual spectra using small spectral windows without strong spectral lines. The obtained values are typically in the range 0.003--0.008, in units of the continuum level. 
To evaluate the presence of non-random noise, we convolved the residual spectra with boxy kernels of increasing size ($n$ pixels)  
and analysed the behaviour of the RMS as a function of $n$.
Pixels-to-pixel random noise is expected to decrease as $n^{-1/2}$ while low-frequency noise
is not damped until the use of  wide enough kernels.
In this way, by modelling the observed function RMS($n$) with an $n^{-1/2}$ term plus a low-frequency component, we 
obtained rough estimates of the contribution of non-random noise.
 
Typically random noise (mainly photon noise) is 0.6\% of the continuum level while non--random errors on scales similar to spectral features (10--40 pixels) are of the order of 0.1\%.
Besides that, run--to--run low--frequency modulations (normalization) of the order 0.5\% are present, but they are easily filtered without altering most spectral features. 
In sum, in a typical individual spectrum our detection limit for variations is about 0.6\%, which for most spectral lines, whose profiles cover 10--40 pixels, corresponds to equivalent widths (W$_\mathrm{eq}$) variations of 0.2--0.4 m\AA~\footnote{For weak lines the uncertainty in W$_\mathrm{eq}$ is $\sigma_\mathrm{Weq}=n^{1/2}\cdot\Delta\cdot(S/N)^{-1}$, where $\Delta$ is the pixel size in wavelength units and $n$ the integration aperture in pixels.}.
At this level there is no evident spectral variations other than the resonance doublet lines of \ion{Ca}{ii}. 
On the other hand, even though many spectra of the same run can be combined to improve the S/N ratio, we will consider the flux uncertainty to be always at least 0.1\% on the continuum level.

\begin{figure}[!ht]
\centering
\includegraphics[width=8cm, angle =0]{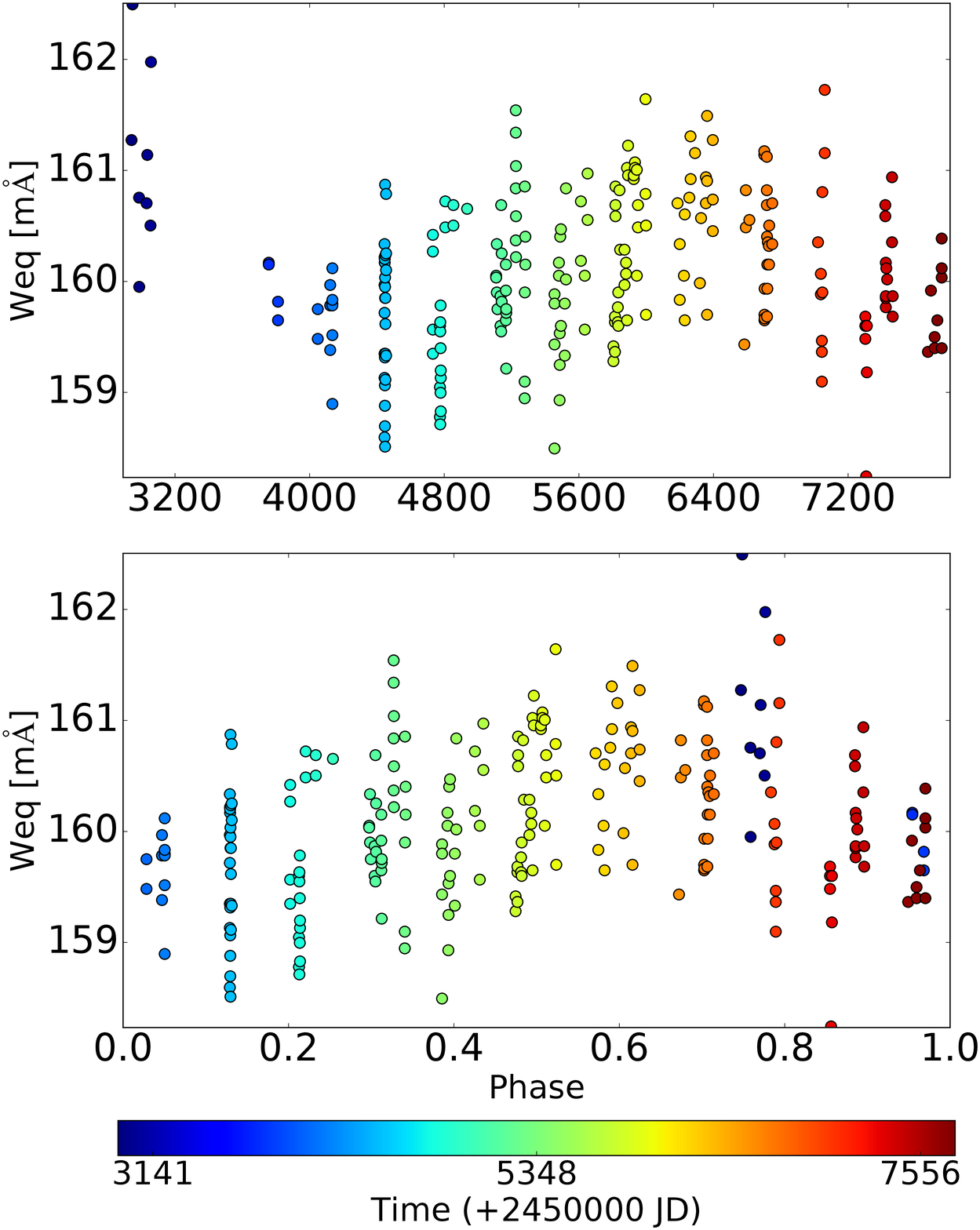}
\caption{Upper panel: W$_\mathrm{eq}$ variation of \ion{Fe}{ii} 5018 {\AA} line. Lower panel: W$_\mathrm{eq}$ phased with a 10.8 yr period.}
\label{fig:m05018}
\end{figure}

\begin{figure}
\centering
\includegraphics[width=\hsize]{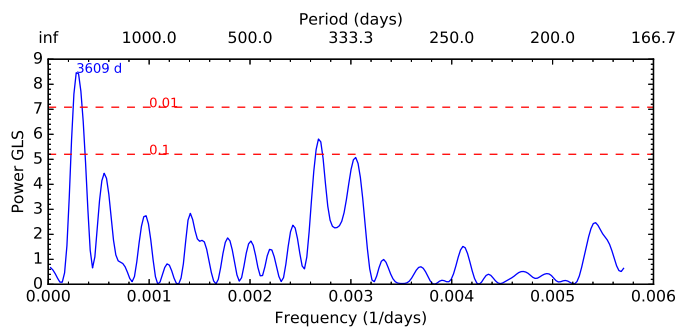}
\caption{GLS periodogram of the W$_\mathrm{eq}$ data plotted in Figure \ref{fig:m05018}.}
\label{fig:m05018b}
\end{figure}

As a first strategy to search for intensity variation of metallic lines, we calculate the  W$_\mathrm{eq}$ of the line profiles for several strong \ion{Fe}{ii} lines. Figure \ref{fig:m05018} shows the W$_\mathrm{eq}$ variations of the \ion{Fe}{ii} line at 5018 {\AA}.
The point distribution suggests an increase of this \ion{Fe}{ii} line with stellar activity.
The period detection from these data using the GLS periodogram (see Fig.~\ref{fig:m05018b}) should be considered as marginal (FAP $\sim$1.6 $\times 10^{-3}$), according to our fixed cut-off. Thus, the activity cycle is not easily detected in these lines.

With the aim to detect low-level spectral variations related specifically with the time-scale
of the activity cycle, we applied two strategies. 

On the one hand, we calculate two mean spectra at low and high activity.
The high-activity spectrum was calculated by averaging the 42 spectra taken  between  December 2011  and February 2014 (HJD 2,455,926--2,456,705), while the low-activity one was the average of 91 spectra between November 2006 and March 2010 (HJD 2,454,049--2,455,283). Then, we calculated  the difference between them and search for significant features on the difference spectrum. Random errors in these mean spectra are expected to be reduced to about 0.09\% and 0.06\%, respectively. 
Thus considering that spectral features are about 10--20 pixels wide, false variability features of 0.03\% are expected to be present due to random noise.     
This value is smaller than non-random errors mentioned above, which might reach $\sim$0.1\%. We therefore consider spectral lines to be variable if there is a difference of at least 0.2\% between the epochs of low and high activity.

\begin{figure*}[!ht]
\centering
\includegraphics[width=10cm, angle =-90]{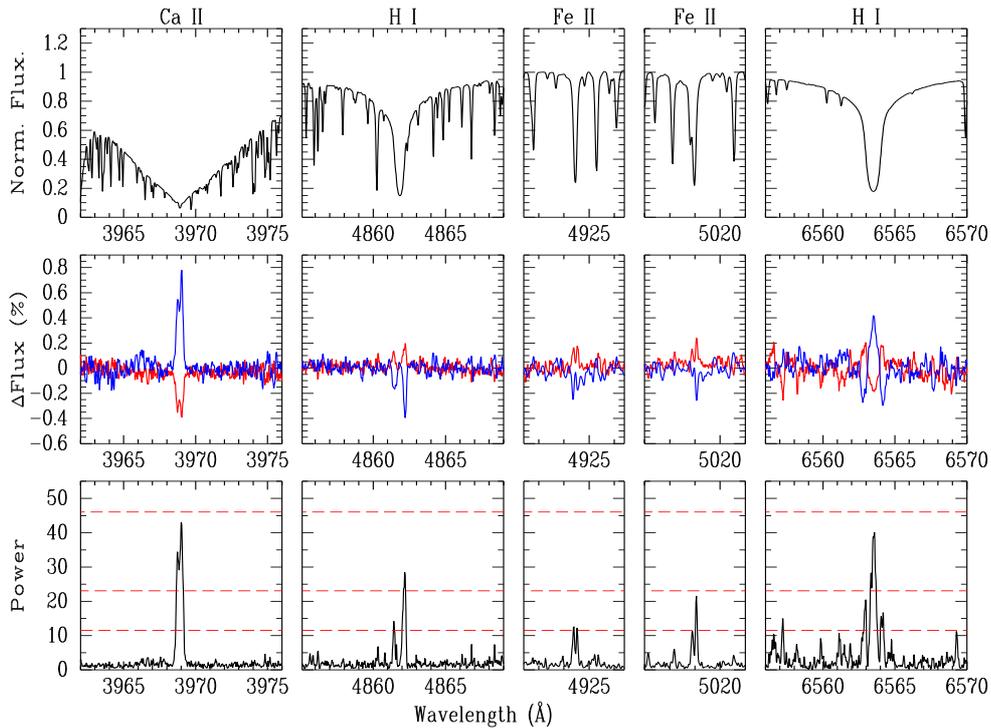}
\caption{Spectral variation of optical lines. Upper panel: Mean spectra near the lines \ion{Ca}{ii} H, H$\beta$, \ion{Fe}{ii} 4924 {\AA}, \ion{Fe}{ii} 5018 {\AA}, and H$\alpha$. Middle panel: Differences between the high-activity (blue line) mean spectrum and the mean reference spectrum and the low-activity (red solid line) mean spectrum and the mean reference spectrum. Lower panel: Periodogram height at $P=3940$ d for each spectrum pixel. Horizontal dashed lines correspond to FAP = 10$^{-5}$, 10$^{-10}$, and 10$^{-15}$ (see text for explanation).}
\label{fig:low-hi}
\end{figure*}

Figure~\ref{fig:low-hi} shows in blue (red) the difference between the high-activity (low-activity) mean spectrum and the mean reference spectrum, around five spectral lines: $\lambda$3968 \ion{Ca}{ii}, H$\alpha$, H$\beta$, and two strong iron lines that were reported to be sensitive to chromospheric activity in HD\,45184. Clear variations are seen in the core of \ion{H}{i} lines. In the position of \ion{Fe}{ii} lines there would be  small differences between low and high activity spectra, although they should be considered as  marginal detections, since they are close to 0.2\%. This is consistent with results obtained from equivalent widths.
 
On the other hand, from the spectral time-series we extracted the light curve corresponding to each wavelength, and calculated the power of the periodogram for the adopted value for the activity period. 
This is shown in the lower panels of Fig.~\ref{fig:low-hi}.
Balmer lines appears  clearly variable with FAP in the range 10$^{-10}$--10$^{-20}$ while iron lines
 present FAP somewhat lower but still over 10$^{-5}$. Note that, being the aim to detect
spectral features varying with the activity cycle, these FAP calculations corresponds to a fixed period 
given a priori, and not to the maximum peak for a grid of frequencies.

\begin{figure}[!ht]
\centering
\includegraphics[width=8cm, angle =0]{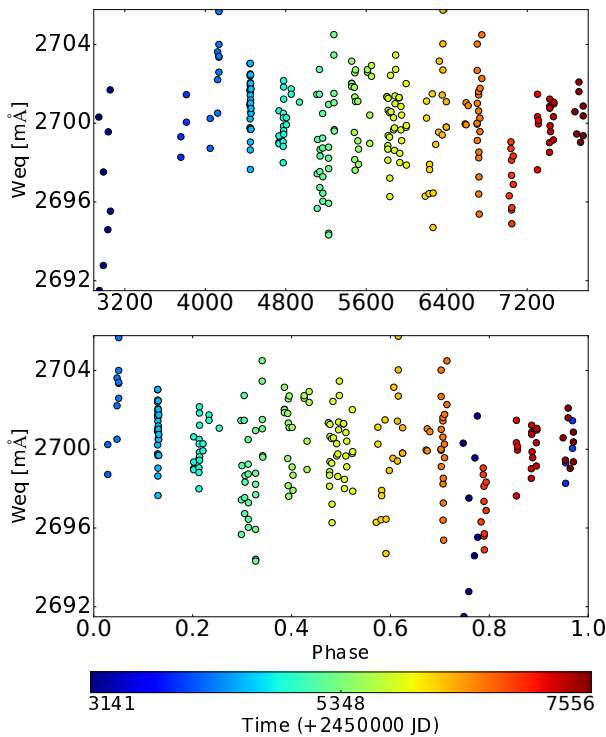}
\caption{Upper panel: W$_\mathrm{eq}$ variation of H$\alpha$ {\AA} line. Lower panel: W$_\mathrm{eq}$ phased with a 10.8 yr period.}
\label{fig:weq6562}
\end{figure}

\begin{figure}
\centering
\includegraphics[width=\hsize]{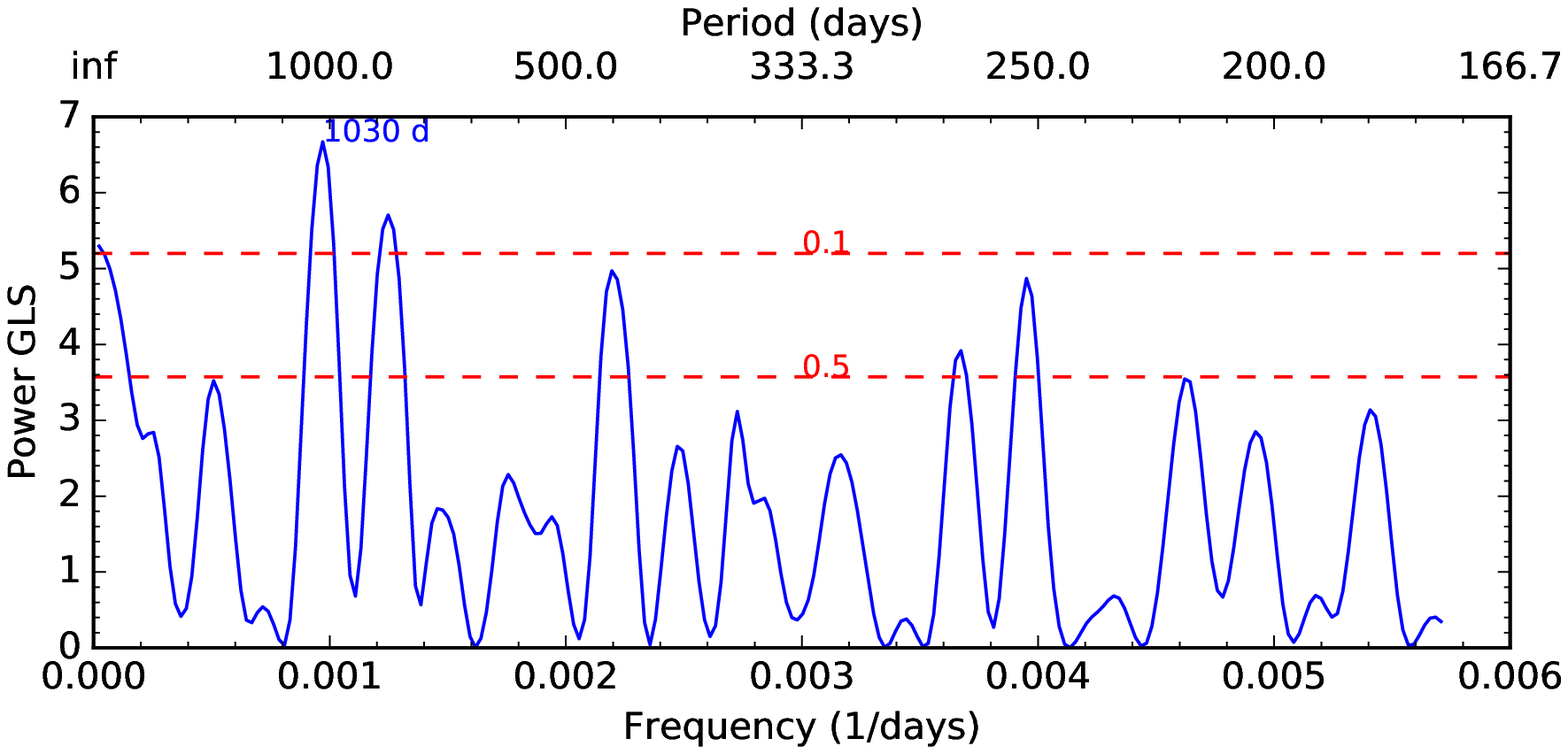}
\caption{GLS periodogram of the W$_\mathrm{eq}$ data plotted in Figure \ref{fig:weq6562}.}
\label{fig:weq6562gls}
\end{figure}

Besides \ion{Ca}{ii} lines, the line most clearly variable is H$\alpha$. The variations cover about 2--3 \AA \ around the core of the line, making the detection highly reliable. The variation affects mainly the shape of the line profile, without significant variations in W$_{eq}$, since the flux variations at the very center of the line is opposite to that at about 0.7 \AA \ from the line center. This kind of profile variations go unnoticed to detection techniques based on flux integration in a given window. To show this fact, we have plotted in Figure \ref{fig:weq6562} the time series for the W$_\mathrm{eq}$ of H$\alpha$ line, while the corresponding GLS periodogram is showed in Figure \ref{fig:weq6562gls}. As can be observed in these figures, there is no significant periodic signal in the equivalent width time series.

\begin{figure}[!ht]
\centering
\includegraphics[width=5cm, angle =-90]{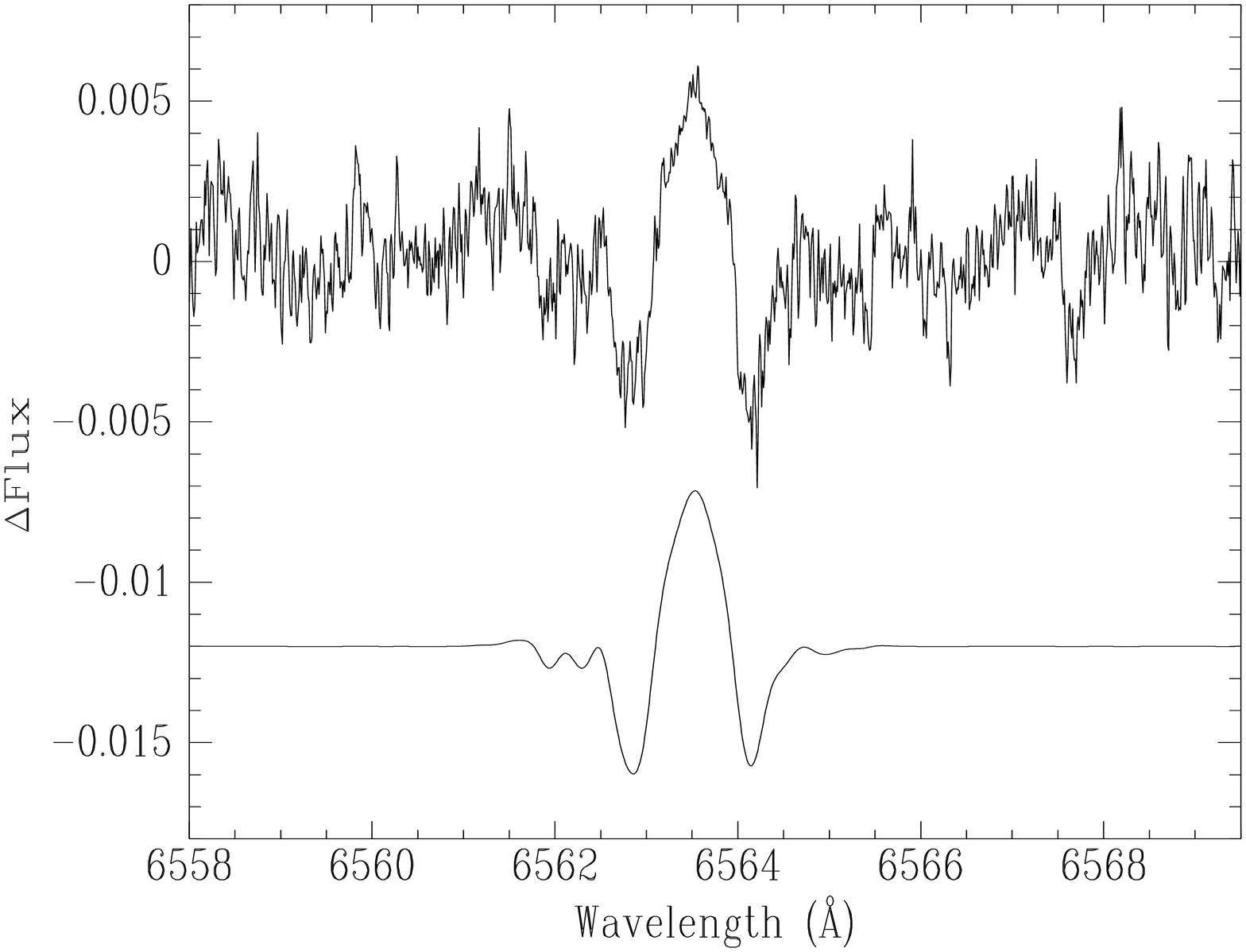}
\caption{Difference between the high- and low-activity mean spectra (up) and the corresponding H$\alpha$ mask (bottom) used to analyse the spectral variations. Flux is normalized to the original continuum level; the lower spectrum has been shifted for better visibility.}\label{fig:mask}
\end{figure}

A similar behaviour was found in HD\,45184, where we used the kurtosis of the line core to analyse the variability. In the present paper we adopted a different strategy. To use the shape of H$\alpha$ as activity proxy, we made a mask representing the activity effects on the profile of H$\alpha$. Specifically, we calculated the difference between the high and low activity spectra and applied a smoothing by convolving  the profile with a Gaussian of width $\sigma=0.2$ \AA~ and apodized the edges by multiplying by a cosine with full intensity in the central 1.5 \AA~ and zero out of a window of 5.0 \AA~ (See Fig.~\ref{fig:mask}).

We define an activity index $I_\alpha$ as the integral of the product of this mask $M$ with the observed spectrum $S$: 
\begin{equation}
I_{\alpha} = \int S(\lambda)\cdot M(\lambda) d\lambda
\label{eq.maskone}
\end{equation}

where $M=H-L$, being $H$ the high-activity mean spectrum and $L$ the low-activity mean spectrum.
Note that $I_\alpha$ is positive when the spectrum $S$ is more similar to $H$ than to $L$ and vice versa: 

\begin{equation}
I_\alpha = \int (S-L)^2d\lambda - \int (S-H)^2d\lambda + \mathrm{const.}
\label{eq.masktwo}
\end{equation}

\begin{figure}[!ht]
\centering
\includegraphics[width=7cm]{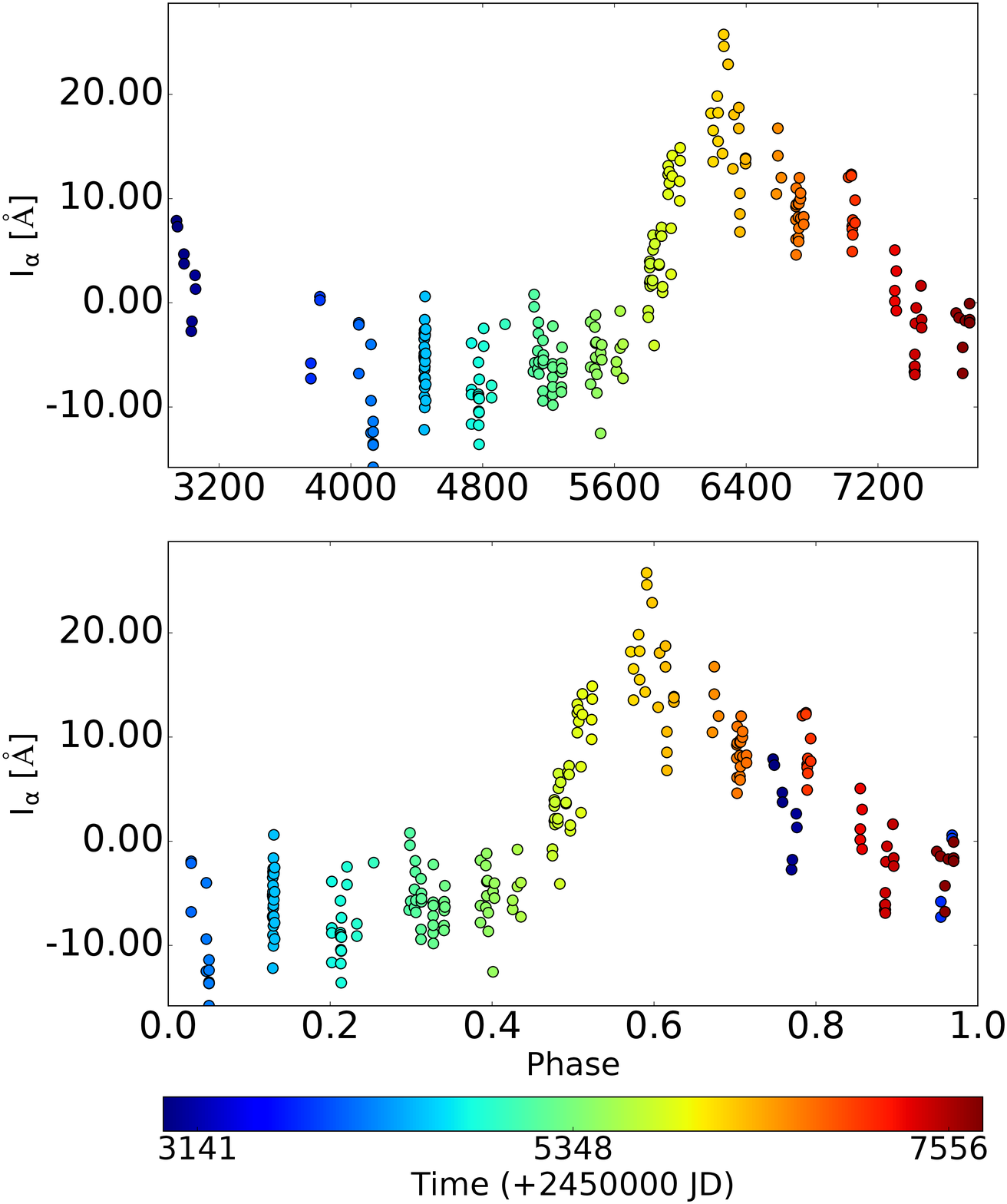}
\caption{Upper panel: Time series of the $I_{\alpha}$ activity parameter. Lower panel: $I_{\alpha}$ variation phased with a $\sim$ 10.8 yr period.}
\label{fig:halpha}
\end{figure}

We measured the activity parameter $I_\alpha$ in all HARPS spectra of HD\,38858 obtaining 
the results shown in Figure \ref{fig:halpha} (upper panel). The spectral variations clearly replicate the behaviour of the Mt Wilson activity index. Also, 
in an attempt to verify the period of the activity cycle, in Figure \ref{fig:halpha} (lower panel) 
we show the $I_{\alpha}$ parameter phased with the chromospheric $\sim$10.8 yr activity period. 
It can be observed a good agreement between both the \ion{Ca}{ii} and $I_{\alpha}$ variations (see Figure \ref{plot.0} for comparison).

\section{Discussion and conclusions}

In this work we perform a detailed study of stellar activity in the solar-analogue HD\,38858, which 
shows several spectral lines with signs of appreciable variations. To do so, we have analysed a total of 242 HARPS and CASLEO spectra. 
As a result, we found a chromospheric activity cycle of $\sim$10.8 yr with an amplitude $\Delta S$ $\sim$ 0.013, a mean S$_{MW}$ index of 0.170 (log($\mathrm{R}'_\mathrm{HK}$) = -4.94), and a shape of the stellar cycle similar to the solar one. Although the solar cycle fluctuates in amplitude, shape, and length \citep{2010LRSP....7....3C,2010LRSP....7....1H}, the corresponding values obtained for HD\,38858 indicate that its activity behaviour could be similar to the Sun.

On the other hand, the GLS periodogram of HARPS radial velocities reveals the presence of two significant peaks. One of them corresponding to the long-term chromospheric activity cycle, which is produced by the active regions situated in the chromosphere of the star. The second peak of 420 $\pm$ 4 d is an alias of the more significant peak, which was initially interpreted as a super-Earth planet by \citet{2011arXiv1109.2497M}. However, our analysis clearly shows that this radial velocity modulation is produced by the stellar activity, in agreement with the previous claim of \citet{2015MNRAS.449.3121K}.

It is believed that there is no a simple relation between different activity proxies. For instance, for the case of the 
\ion{Ca}{ii} H\&K and H$\alpha$ line correlation in solar type stars, some authors attribute the discrepancies to the activity level and metallicity \citep{2014A&A...566A..66G}, while others suggest that this correlation is complex and depends on the star under consideration \citep{2007A&A...469..309C}.
In addition, recently \citet{2016A&A...589A.135F} found in HD\,45184, a star with similar physical properties to HD\,38858, that the statistical
moments, including the W$_\mathrm{eq}$ of Balmer and some \ion{Fe}{ii} lines (4924 {\AA}, 5018 {\AA}, and 5169 {\AA}), are modulated by the chromospheric cycle.
The existence of another activity proxy different from the \ion{Ca}{ii} H\&K lines is important not only as a tool for searching activity cycles, but also to distinguish planetary from stellar signals in those planet-search radial-velocity surveys where \ion{Ca}{ii} H\&K lines are not included. Here, the \ion{Ca}{ii} H\&K modulation (chromospheric cycle) of HD 38858 is also evident in Balmer lines (H$\alpha$ and H$\beta$), while \ion{Fe}{ii} at 4924 {\AA} and  5018 {\AA},  result in marginal detections.

\citet{2014A&A...566A..66G} analysed the possible correlation between log($\mathrm{R}'_\mathrm{HK}$) and log($I_{H\alpha}$)\footnote{This index measures the flux in the core of the H$\alpha$ line \citep[see][for details]{2014A&A...566A..66G}.} in a stellar sample of 271 FGK stars through the Pearson correlation coefficient. As a result, only 23\% out of these stars were correlated (20\% showing a positive correlation and 3\% showing a negative one). Notably, they included HD 38858 in the group of stars that did not show any correlation, with a high level of significance. In the present paper we detected in HD38858 spectral variations affecting the shape of the core of H$\alpha$, and demonstrated that these variations do correlate with the activity index log($\mathrm{R}'_\mathrm{HK}$). This finding substantiates our previous results obtained for HD 45184, another star without H$\alpha$-log($\mathrm{R}'_\mathrm{HK}$) correlation  according to \citet{2014A&A...566A..66G}, for which we had detected variations in the H$\alpha$ profile through the analysis of the kurtosis of the line core. 
Therefore, our results suggest that, at least in solar-analogues stars, the shape of the line profile of H$\alpha$ would be a better candidate than net flux if we want to search for correlations with the stellar activity.

This has relevance in the definition of alternative activity proxies. \citet{2014A&A...566A..66G}  also demonstrate that the flux at the H$\alpha$ core is not a sensitive indicator for detecting magnetic cycles. In  a sample of 271 stars they were able to find activity cycles in the 26\% using the log($\mathrm{R}'_\mathrm{HK}$) index, while only in the 3\% when using their $I_{H\alpha}$ index. However, we detected chromospheric cycles in both stars HD 45184 and HD 38858 (with periods of 5.14 and 10.8 yr, respectively) studying the H$\alpha$ variations. Then, both HD 45184 and HD 38858 are remarkable objects, showing clear cycles detected in log($\mathrm{R}'_\mathrm{HK}$) and also in the H$\alpha$ profile, with a supposed null $I_{H\alpha}$  correlation \citep{2014A&A...566A..66G}.

In the present state of knowledge, gathering observational data is crucial. The only way to shed light on these issues is
to perform detailed spectral analyses of the different activity proxies in a wide sample of solar type stars.
For this reason, we are currently starting to search for \ion{Fe}{ii}
line variations and study the \ion{Ca}{ii} H\&K and H$\alpha$ correlation in a
significant star sample using available HARPS data.

\begin{acknowledgements}
We warmly thank the anonymous referee for the careful reading of the manuscript and constructive comments that enabled the work to be improved. Also, M.F., M.J and F.M.L. acknowledge the financial support from CONICET in the forms of Post-Doctoral Fellowships.

\end{acknowledgements}

\bibliographystyle{aa}
\bibliography{references}
\end{document}